\documentclass[aps,preprint]{revtex4}%
\usepackage{amsfonts}
\usepackage{amsmath}
\usepackage{amssymb}
\usepackage{graphicx}%
\providecommand{\U}[1]{\protect \rule{.1in}{.1in}}

\begin{document}
\title[FFLO in 3D Fermi gas with phase fluctuations]{Do phase fluctuations influence the Fulde-Ferrell-Larkin-Ovchinnikov state in
a 3D\ Fermi gas?}
\author{Jeroen P.A. Devreese$^{1,2}$}
\email{jeroen.devreese@uantwerpen.be}
\author{Jacques Tempere$^{1,3}$}
\affiliation{$^{1}$TQC (Theory of Quantum systems and Complex systems), Universiteit
Antwerpen, B-2020 Antwerpen, Belgium}
\affiliation{$^{2}$School of Physics, Georgia Institute of Technology, Atlanta, GA\ 30332, USA}
\affiliation{$^{3}$Lyman Laboratory of Physics, Harvard University, Cambridge, MA 02138, USA.}

\begin{abstract}
In ultracold Fermi gases, the effect of spin-imbalance on superfluidity has
been the subject of intense study. One of the reasons for this is that
spin-imbalance frustrates the Bardeen-Cooper-Schrieffer (BCS) superfluid
pairing mechanism, in which fermions in different spin states combine into
Cooper pairs with zero momentum. In 1964, it was proposed that an exotic
superfluid state called the Fulde-Ferrell-Larkin-Ovchinnikov (FFLO) state, in
which the Cooper pairs have nonzero momentum, could exist in a spin-imbalanced
Fermi gas. At the saddle-point (mean field) level, it has been shown that the
FFLO state only occupies a very small sliver in the ground state phase diagram
of a 3D Fermi gas. However, a question that remains to be investigated is:
what is the influence of phase fluctuations around the saddle point on the
FFLO state? In this work we show that phase fluctuations only lead to
relatively small quantitative corrections to the presence of the FFLO state in
the saddle-point phase diagram of a 3D spin-imbalanced Fermi gas. Starting
from the partition function of the system, we calculate the effective action
within the path-integral adiabatic approximation. The action is then expanded
up to second order in the fluctuation field around the saddle point, leading
to the fluctuation free energy. Using this free energy, we calculate
corrections due to phase fluctuations to the BCS-FFLO transition in the
saddle-point phase diagram. At temperatures at which the FFLO state exists, we
find only small corrections to the size of the FFLO area. Our results suggest
that fluctuations of the phase of the FFLO order parameter, which can be
interpreted as an oscillation of its momentum vector, do not cause an
instability of the FFLO state with respect to the BCS state.

\end{abstract}
\date{\today}

\maketitle

\section{Introduction}

In the last two decades, ultracold quantum gases have been the subject of many
theoretical and experimental investigations \cite{Reviews cold atoms}. Among
the many systems that have been studied, ultracold Fermi gases have received
wide attention \cite{Reviews Fermi gases}. Due to the experimental
controllability achieved with ultracold gases, quantum many-body phenomena
such as fermionic superfluidity can be studied in great detail in these
systems. By controlling and tuning the interaction strength between fermions
in different states using Feshbach resonances \cite{Feshbach resonance}, it
has become possible to study the crossover from a Bardeen-Cooper-Schrieffer
(BCS) superfluid state of weakly interacting Cooper pairs, to a Bose-Einstein
condensate (BEC) of strongly coupled molecules \cite{BCS-BEC crossover
(1),BCS-BEC crossover (2)}.

Aside from the interaction strength, another important parameter that can be
tuned is the population imbalance between fermions in different states. This
parameter is of importance because spin-imbalance frustrates the BCS
superfluid pairing mechanism. In the BCS state, pairing between fermions
occurs at the Fermi surface. However, a population imbalance will create a gap
between the\ Fermi surfaces of the two spin states, making the BCS state
energetically less favorable. Theoretically it was predicted that at a certain
critical spin-imbalance, known as the Clogston-Chandrasekhar limit
\cite{Clogston-Chandrasekhar}, a first order phase transition from the BCS
state to the normal state would occur. By preparing a Fermi gas in one
hyperfine state, and using a radio-frequency sweep to create a mixture of two
hyperfine states (labeled spin-up and spin-down), the transition from a
superfluid to a normal gas, induced by spin-imbalance, was demonstrated
experimentally \cite{Superfluid to normal transition}.

At this point, the question remains whether a non-uniform superfluid can exist
in a spin-imbalanced 3D Fermi gas. The most prominent example of non-uniform
superfluidity is the Fulde-Ferrell-Larkin-Ovchinnikov (FFLO) state, which was
proposed independently by Fulde and Ferrell (FF) \cite{Fulde-Ferrell} and by
Larkin and Ovchinnikov (LO) \cite{Larkin-Ovchinnikov} in 1964. The FFLO state
differs from the BCS state in that it has Cooper pairs with non-zero momentum,
which in position space results in an oscillating superfluid order parameter.
It was suggested that this exotic superfluid state could exist at non-zero
polarization. In part of the literature, a further distinction is made between
the FF\ state and the LO state: the former has one momentum component, whereas
the latter is the superposition of two momentum components of equal magnitude
but opposite sign. In this paper, we will focus on the FF\ state but we will
henceforth call this the FFLO state, bearing in mind that we mean the
superfluid state with one momentum component.

Following the success of creating a spin-imbalanced Fermi gas, the theoretical
investigation of the FFLO superfluid state was intensified. The first studies
focused on the three-dimensional (3D) Fermi gas, at the saddle-point
(mean-field) level \cite{FFLO 3D}, and found that the FFLO state is only
present in a very small sliver of the ground-state phase diagram \cite{FFLO 3D
FD}. The 1D case has also received wide theoretical attention, and has proven
to be a promising setup for detecting the FFLO state. In 1D, the presence of
the FFLO state in the ground-state phase diagram is much larger compared to
the 3D case \cite{FFLO 1D}. Following these theoretical predictions, the first
indirect experimental evidence for the FFLO state was found in a 1D Fermi gas
by the Hulet group at Rice University \cite{Hulet FFLO 1D}. Inspired by this
success, several new experimental detection techniques have been proposed
\cite{Detection techniques}, both for the 1D and for the 3D case. However, in
the latter case, the FFLO state still eludes experimental observation.

To acquire a better understanding of the FFLO state in a 3D Fermi gas, it is
necessary to go beyond the mean-field level, which, while resulting in
quantitatively correct results in the limit of weak interaction (BCS limit) at
temperature zero, offers a qualitative description at best for temperatures
above zero or for stronger interactions. Up till now, little attention has
been devoted to this subject and to the effect of fluctuations on the FFLO
state in general. One important exception is the work by Radzihovsky,
\cite{Radzihovsky} in which a low-energy model for the Fulde-Ferrell state and
for the Larkin-Ovchinnikov state is developed, with an in-depth focus on the
nature of the emerging Goldstone modes for the latter state.

In this paper, we contribute to this subject by explicitly studying the effect
of phase fluctuations on the presence of the FFLO state in the phase diagram
of a 3D Fermi gas. Our main motivation is the following: the FFLO state is
characterized by a momentum component $\mathbf{Q}$, which means that the
rotational symmetry of the system is spontaneously broken by this state. The
momentum component $\mathbf{Q}$ results in an oscillating phase of the order
parameter in position space. Because of this, fluctuations of the phase of the
order parameter are equivalent to fluctuations in the direction of the
momentum $\mathbf{Q}$. Since a 3D Fermi gas exhibits spherical symmetry, these
fluctuations cost zero energy. Hence, one would expect these fluctuations to
proliferate and destabilize the FFLO state. Our main point of interest is to
see whether the region of FFLO in the phase diagram of a 3D Fermi gas vanishes
due to phase fluctuations, which would help to understand why this state has
not been observed experimentally in 3D. To the best of our knowledge, this
specific problem has not yet been studied in literature.

The rest of this paper is organized as follows. In section
\ref{Hydrodynamic effective action} we derive a hydrodynamic effective action,
starting from the partition function of a 3D\ Fermi gas with spin-imbalance,
within the path-integral adiabatic approximation. In section
\ref{Fluctuation action} we perform an expansion of the action up to second
order in the fluctuation field, which leads to the fluctuation part of the
action. From the fluctuation action, the fluctuation free energy is readily
derived. Subsequently, in section \ref{Phase diagram}, using this free energy,
we calculate the phase diagram of the system and determine whether corrections
emerge by taking into account phase fluctuations. Finally in section
\ref{Conclusions} we draw conclusions.

\section{Calculating the hydrodynamic effective action
\label{Hydrodynamic effective action}}

In this section, the effective action describing the FFLO state in a
spin-imbalanced 3D Fermi gas, with the inclusion of phase fluctuations around
the saddle point, is calculated within a hydrodynamic approach. The starting
point of this derivation is the partition function of the system, written as a
path integral over fermionic Grassmann fields $\bar{\psi}_{\mathbf{x}%
,\tau,\sigma}$ and $\psi_{\mathbf{x},\tau,\sigma}$:
\begin{equation}
\mathcal{Z}=\int \mathcal{D}\bar{\psi}_{\mathbf{x},\tau,\sigma}\mathcal{D}%
\psi_{\mathbf{x},\tau,\sigma}~e^{-S\left(  \bar{\psi}_{\mathbf{x},\tau,\sigma
},\psi_{\mathbf{x},\tau,\sigma}\right)  },
\end{equation}
where the action consists of a single-particle term and an interaction term:%
\begin{align}
S\left(  \bar{\psi}_{\mathbf{x},\tau,\sigma},\psi_{\mathbf{x},\tau,\sigma
}\right)   &  =\int_{0}^{\beta}d\tau \int d\mathbf{x}\sum_{\sigma}\bar{\psi
}_{\mathbf{x},\tau,\sigma}\left(  \frac{\partial}{\partial \tau}-\nabla
_{\mathbf{x}}^{2}-\mu_{\sigma}\right)  \psi_{\mathbf{x},\tau,\sigma
}\nonumber \\
&  +\int_{0}^{\beta}d\tau \int d\mathbf{x\int}d\mathbf{y~}\bar{\psi
}_{\mathbf{x},\tau,\uparrow}\bar{\psi}_{\mathbf{y},\tau,\downarrow}V\left(
\mathbf{x-y}\right)  \psi_{\mathbf{y},\tau,\downarrow}\psi_{\mathbf{x}%
,\tau,\uparrow}. \label{Action}%
\end{align}
The action (\ref{Action}) is written in imaginary time $\tau=it$, $\mathbf{x}$
and $\mathbf{y}$ represent 3D position vectors, $\beta=1/k_{B}T$ is the
inverse temperature and $\mu_{\sigma}$ is the chemical potential of the
fermions in the spin state $\sigma=\left \{  \uparrow,\downarrow \right \}  $.
Furthermore, we use $\hbar=2m=1$ as units. In the interaction term, $V\left(
\mathbf{x-y}\right)  $ represents a general interparticle potential. In this
paper, only s-wave scattering will be considered at ultracold temperatures,
therefore the interaction potential will be replaced by a pseudo-potential:%
\begin{equation}
V\left(  \mathbf{x-y}\right)  =g\delta \left(  \mathbf{x-y}\right)  ,
\label{Contact potential}%
\end{equation}
where the interaction strength $g$ is related to the s-wave scattering length
$a_{s}$ through \cite{Pseudopotential,Pseudopotential (2)}:%
\begin{equation}
\frac{1}{g}=\frac{1}{8\pi a_{s}}-\int \frac{d\mathbf{k}}{\left(  2\pi \right)
^{3}}\frac{1}{2k^{2}}. \label{1/g(as) (2)}%
\end{equation}
An alternative, elegant derivation of expression (\ref{1/g(as) (2)}) can be
found in \cite{Castin}.

Using the standard Hubbard-Stratonovich transformation, the fourth-degree
interaction term in (\ref{Action}) is rewritten as a sum of two second-degree
terms. The cost of this transformation is that an additional path integral,
over auxiliary bosonic fields $\Delta_{\mathbf{x},\tau}$ and $\bar{\Delta
}_{\mathbf{x},\tau}$, is introduced. These fields are physically relevant, as
they are interpreted as the fields of the fermion pairs. Following the
Hubbard-Stratonovich transformation, the partition function becomes%
\begin{align}
\mathcal{Z}  &  =\int \mathcal{D}\bar{\psi}_{\mathbf{x},\tau,\sigma}%
\mathcal{D}\psi_{\mathbf{x},\tau,\sigma}\int \mathcal{D}\bar{\Delta
}_{\mathbf{x},\tau}\mathcal{D}\Delta_{\mathbf{x},\tau}\nonumber \\
&  \times \exp \left[  -\int_{0}^{\beta}d\tau \int d\mathbf{x}\sum_{\sigma}%
\bar{\psi}_{\mathbf{x},\tau,\sigma}\left(  \frac{\partial}{\partial \tau
}-\nabla_{\mathbf{x}}^{2}-\mu_{\sigma}\right)  \psi_{\mathbf{x},\tau,\sigma
}\right. \nonumber \\
&  \left.  +\int_{0}^{\beta}d\tau \int d\mathbf{x}\left(  \frac{\bar{\Delta
}_{\mathbf{x},\tau}\Delta_{\mathbf{x},\tau}}{g}-\Delta_{\mathbf{x},\tau}%
\bar{\psi}_{\mathbf{x},\tau,\uparrow}\bar{\psi}_{\mathbf{x},\tau,\downarrow
}-\bar{\Delta}_{\mathbf{x},\tau}\psi_{\mathbf{x},\tau,\downarrow}%
\psi_{\mathbf{x},\tau,\uparrow}\right)  \right]  . \label{Partition function}%
\end{align}
One way to introduce fluctuations into the partition function is to write the
bosonic fields as the sum of a saddle-point contribution and a fluctuation
contribution: $\Delta_{\mathbf{x},\tau}=\Delta_{\mathbf{x},\tau}^{\left(
sp\right)  }+\phi_{_{\mathbf{x},\tau}}$ and $\bar{\Delta}_{\mathbf{x},\tau
}=\Delta_{\mathbf{x},\tau}^{\left(  sp\right)  }+\bar{\phi}_{_{\mathbf{x}%
,\tau}}$. In this way, amplitude- and phase fluctuations are fully
intertwined, and both are automatically taken into account \cite{Nozieres
Schmitt-Rink}. An alternative way, which will be used in this paper since it
allows (at a later stage in this calculation) to focus solely on phase
fluctuations, is to write the bosonic fields in terms of an amplitude and a
phase
\begin{equation}
\left \{
\begin{array}
[c]{l}%
\Delta_{\mathbf{x},\tau}=\left \vert \Delta_{\mathbf{x},\tau}\right \vert
e^{i\theta_{\mathbf{x},\tau}}\\
\bar{\Delta}_{\mathbf{x},\tau}=\left \vert \Delta_{\mathbf{x},\tau}\right \vert
e^{-i\theta_{\mathbf{x},\tau}}%
\end{array}
\right.  , \label{Amplitude and phase}%
\end{equation}
where both $\left \vert \Delta_{\mathbf{x},\tau}\right \vert $ and
$\theta_{\mathbf{x},\tau}$ are real fields. Upon substituting
(\ref{Amplitude and phase}) into the partition function
(\ref{Partition function}), it is convenient to apply the following gauge
transformation to the fermionic fields: $\psi_{\mathbf{x},\tau,\sigma
}\rightarrow \psi_{\mathbf{x},\tau,\sigma}e^{i\theta_{\mathbf{x},\tau}/2}$.
After working out the derivatives in (\ref{Partition function}), the partition
function becomes%
\begin{align}
\mathcal{Z}  &  =\int \mathcal{D}\bar{\psi}_{\mathbf{x},\tau,\sigma}%
\mathcal{D}\psi_{\mathbf{x},\tau,\sigma}\int \mathcal{D}\left \vert
\Delta_{\mathbf{x},\tau}\right \vert \mathcal{D}\theta_{\mathbf{x},\tau
}\nonumber \\
&  \times \exp \left(  -\int_{0}^{\beta}d\tau \int d\mathbf{x~}\bar{\eta
}_{\mathbf{x},\tau}\left(  -\mathbb{G}_{\mathbf{x},\tau}^{-1}\right)
\eta_{\mathbf{x},\tau}+\int_{0}^{\beta}d\tau \int d\mathbf{x}\frac{\left \vert
\Delta_{\mathbf{x},\tau}\right \vert ^{2}}{g}\right)  ,
\label{Partition function (1)}%
\end{align}
where the following Nambu spinors were used:%
\begin{equation}
\bar{\eta}_{\mathbf{x},\tau}=\left(
\begin{array}
[c]{cc}%
\bar{\psi}_{\mathbf{x},\tau,\uparrow} & \psi_{\mathbf{x},\tau,\downarrow}%
\end{array}
\right)  \text{ and }\eta_{\mathbf{x},\tau}=\left(
\begin{array}
[c]{c}%
\psi_{\mathbf{x},\tau,\uparrow}\\
\bar{\psi}_{\mathbf{x},\tau,\downarrow}%
\end{array}
\right)  , \label{Nambu spinor}%
\end{equation}
in order to write the inverse Green's function in block-diagonal form, where
the diagonal elements are given by a set of $2\times2$ matrices given by%
\begin{align}
-\mathbb{G}_{\mathbf{x},\tau}^{-1}  &  =\left(  \dfrac{\partial}{\partial \tau
}-\zeta-i\nabla_{\mathbf{x}}\left(  \theta_{\mathbf{x},\tau}\right)
.\nabla_{\mathbf{x}}-\dfrac{i}{2}\nabla_{\mathbf{x}}^{2}\left(  \theta
_{\mathbf{x},\tau}\right)  \right)  \sigma_{0}\nonumber \\
&  -\left(  \nabla_{\mathbf{x}}^{2}+\mu-\dfrac{i}{2}\dfrac{\partial
\theta_{\mathbf{x},\tau}}{\partial \tau}-\dfrac{1}{4}\left[  \nabla
_{\mathbf{x}}\left(  \theta_{\mathbf{x},\tau}\right)  \right]  ^{2}\right)
\sigma_{3}+\left \vert \Delta_{\mathbf{x},\tau}\right \vert \sigma_{1}.
\label{-G^-1 x,t}%
\end{align}
Here $\sigma_{0}$,$~\sigma_{1}$ and $\sigma_{3}$ are Pauli matrices, and
furthermore the definitions of the total chemical potential $\mu=\left(
\mu_{\uparrow}+\mu_{\downarrow}\right)  /2$ and the imbalance chemical
potential $\zeta=\left(  \mu_{\uparrow}-\mu_{\downarrow}\right)  /2$ have been introduced.

The path integral over the Bose field $\left \vert \Delta_{\mathbf{x},\tau
}\right \vert $ in (\ref{Partition function (1)}) cannot be calculated exactly
and hence an approximation has to be made. The most basic approximation is to
replace the field $\left \vert \Delta_{\mathbf{x},\tau}\right \vert $ by a
constant $\Delta$: this is the saddle-point approximation. Since in this
article we want to describe phase fluctuations, we improve on this
approximation by also considering fluctuations of the field $\theta
_{\mathbf{x},\tau}$ around its saddle-point value. The specific choice of the
saddle point is such that the FFLO state is included in the current formalism.
The FFLO state is defined by fermionic pairs that have a non-zero momentum
$\mathbf{Q}$. In position space, this is equivalent with an oscillating phase
of the order parameter $\Delta \exp \left(  i\mathbf{Qx}\right)  $. Hence the
choice of the saddle point, including phase fluctuations, is the following:%
\begin{equation}
\left \{
\begin{array}
[c]{l}%
\left \vert \Delta_{\mathbf{x},\tau}\right \vert =\Delta \\
\theta_{\mathbf{x},\tau}=\mathbf{Q.x+}\delta \theta_{\mathbf{x},\tau}%
\end{array}
\right.  , \label{saddle point + phase fluctuations}%
\end{equation}
where $\delta \theta_{\mathbf{x},\tau}$ is the fluctuation field. At this
point, we explicitly choose to neglect amplitude fluctuations and to focus
solely on phase fluctuations. After substitution of
(\ref{saddle point + phase fluctuations}) in the partition function
(\ref{Partition function (1)}), we find:%
\begin{equation}
\mathcal{Z}=\int \mathcal{D}\bar{\psi}_{\mathbf{x},\tau,\sigma}\mathcal{D}%
\psi_{\mathbf{x},\tau,\sigma}\int \mathcal{D}\delta \theta_{\mathbf{x},\tau}%
\exp \left(  -\int_{0}^{\beta}d\tau \int d\mathbf{x~}\bar{\eta}_{\mathbf{x}%
,\tau}\left(  -\mathbb{G}_{\mathbf{x},\tau}^{-1}\right)  \eta_{\mathbf{x}%
,\tau}+\beta V\frac{\Delta^{2}}{g}\right)  . \label{Partitiesom FL}%
\end{equation}
The inverse Green's function in (\ref{Partitiesom FL}) is given by:
\begin{align}
-\mathbb{G}_{\mathbf{x},\tau}^{-1}  &  =\left(  \dfrac{\partial}{\partial \tau
}-\zeta-i\mathbf{Q}.\nabla_{\mathbf{x}}-i\nabla_{\mathbf{x}}\left(
\delta \theta_{\mathbf{x},\tau}\right)  .\nabla_{\mathbf{x}}-\dfrac{i}{2}%
\nabla_{\mathbf{x}}^{2}\left(  \delta \theta_{\mathbf{x},\tau}\right)  \right)
\sigma_{0}\label{-G^-1 x,t (2)}\\
&  -\left(  \nabla_{\mathbf{x}}^{2}+\mu-\dfrac{i}{2}\dfrac{\partial
\delta \theta_{\mathbf{x},\tau}}{\partial \tau}-\dfrac{Q^{2}}{4}-\nabla
_{\mathbf{x}}\left(  \delta \theta_{\mathbf{x},\tau}\right)  .\dfrac
{\mathbf{Q}}{2}~\mathbf{-}\dfrac{1}{4}\left[  \nabla_{\mathbf{x}}\left(
\delta \theta_{\mathbf{x},\tau}\right)  \right]  ^{2}\right)  \sigma_{3}%
+\Delta \sigma_{1}.\nonumber
\end{align}
The partition function (\ref{Partitiesom FL}) still contains two path
integrals: one over the phase fluctuation field and one over the fermionic
fields. To calculate the fermionic path integral, a transformation to
reciprocal space is necessary, because of the derivatives that are present in
the inverse Green's function. However, since the field $\delta \theta
_{\mathbf{x},\tau}$ is a general function of space and time, this will lead to
an infinite number of non-diagonal terms in the action, making the calculation
intractable. As a remedy, the path-integral adiabatic approximation will be
used. This approximation assumes that the bosonic fluctuation field
$\delta \theta_{\mathbf{x},\tau}~$varies slowly in time and space compared to
the fermionic fields $\bar{\psi}_{\mathbf{x},\tau,\sigma}$ and $\psi
_{\mathbf{x},\tau,\sigma}$. As a result, for a given configuration of the
fluctuation field, the configuration of fermionic fields can be coarse-grained
by averaging over the `fast' degrees of freedom:%
\begin{equation}
\bar{\psi}_{\mathbf{x},\tau,\sigma}\psi_{\mathbf{x},\tau,\sigma}%
\rightarrow \frac{1}{\beta V}\int d\tau^{\prime}\int d\mathbf{x}^{\prime}%
\bar{\psi}_{\mathbf{x}^{\prime},\tau^{\prime},\sigma}^{\left(  \mathbf{x}%
,\tau \right)  }\psi_{\mathbf{x}^{\prime},\tau^{\prime},\sigma}^{\left(
\mathbf{x},\tau \right)  }. \label{PIAA}%
\end{equation}
Here $\left(  \mathbf{x},\tau \right)  $ are the space-time points for the
`slow' subsystem, and $\left(  \mathbf{x}^{\prime},\tau^{\prime}\right)  $ are
the space-time points for the `fast' subsystem. The appearance of additional
degrees of freedom is due to the assumption that for boson configurations
$\delta \theta_{\mathbf{x},\tau}$ the averaging over fermion configurations can
be performed while keeping the bosonic field constant. This method was used
successfully for a hydrodynamic description of
the\ Berezinskii-Kosterlitz-Thouless transition in a 2D Fermi gas \cite{Klimin
Tempere Devreese}. Using expression (\ref{PIAA}), the Fourier transformation
of the fermionic fields can be performed independently of the fluctuation
field (see appendix \ref{Fourier transform}). After Fourier transformation,
the path integral over fermionic fields is quadratic and can be calculated
exactly. The partition function then becomes:%
\begin{equation}
\mathcal{Z}=\int \mathcal{D\delta}\theta_{\mathbf{x},\tau}\exp \left(  \frac
{1}{\beta V}\int_{0}^{\beta}d\tau \int d\mathbf{x}\sum_{\mathbf{k},\omega_{n}%
}\log \left(  -\det \left \{  -\mathbb{G}_{\mathbf{k},\omega_{n}}^{-1}\left[
\delta \theta \left(  \mathbf{x},\tau \right)  \right]  \right \}  \right)  +\beta
V\frac{\Delta^{2}}{g}\right)  , \label{Partition Function (2)}%
\end{equation}
with $\mathbf{k}$ the momentum of the fermionic fields and $\omega_{n}=\left(
2n+1\right)  \pi/\beta$ the fermionic Matsubara frequencies. In
(\ref{Partition Function (2)}) the inverse Green's function is given by%
\begin{equation}
-\mathbb{G}_{\mathbf{k},\omega_{n}}^{-1}\left[  \delta \theta \left(
\mathbf{x},\tau \right)  \right]  =\left(
\begin{array}
[c]{cc}%
-i\omega_{n}-\tilde{\zeta}_{\mathbf{k,Q}}^{\left(  \theta \right)  }+\tilde
{\xi}_{\mathbf{k,Q}}^{\left(  \theta \right)  } & \Delta \\
\Delta & -i\omega_{n}-\tilde{\zeta}_{\mathbf{k,Q}}^{\left(  \theta \right)
}-\tilde{\xi}_{\mathbf{k,Q}}^{\left(  \theta \right)  }%
\end{array}
\right)  . \label{Green's function}%
\end{equation}
To write (\ref{Green's function}) in a compact form, the following notations
were introduced:%
\begin{equation}
\left \{
\begin{array}
[c]{l}%
\tilde{\zeta}_{\mathbf{k,Q}}^{\left(  \theta \right)  }=\zeta_{\mathbf{k,Q}%
}+\zeta_{\mathbf{k}}^{\left(  \theta \right)  }\\
\zeta_{\mathbf{k,Q}}=\zeta~-\mathbf{k.Q}\\
\zeta_{\mathbf{k}}^{\left(  \theta \right)  }=\mathbf{-~}\nabla_{\mathbf{x}%
}\left(  \delta \theta_{\mathbf{x},\tau}\right)  .\mathbf{k}+\dfrac{i}{2}%
\nabla_{\mathbf{x}}^{2}\left(  \delta \theta_{\mathbf{x},\tau}\right)
\end{array}
\right.  , \label{zeta}%
\end{equation}
and%
\begin{equation}
\left \{
\begin{array}
[c]{l}%
\tilde{\xi}_{\mathbf{k,Q}}^{\left(  \theta \right)  }=k^{2}-\tilde{\mu
}_{\mathbf{Q}}^{\left(  \theta \right)  }~~\\
\tilde{\mu}_{\mathbf{Q}}^{\left(  \theta \right)  }=\mu_{\mathbf{Q}}%
+\mu_{\mathbf{Q}}^{\left(  \theta \right)  }\\
\mu_{\mathbf{Q}}=\mu-\dfrac{Q^{2}}{4}\\
\mu_{\mathbf{Q}}^{\left(  \theta \right)  }=-\nabla_{\mathbf{x}}\left(
\delta \theta_{\mathbf{x},\tau}\right)  .\dfrac{\mathbf{Q}}{2}-\dfrac{i}%
{2}\dfrac{\partial \delta \theta_{\mathbf{x},\tau}}{\partial \tau}-\dfrac{1}%
{4}\left[  \nabla_{\mathbf{x}}\left(  \delta \theta_{\mathbf{x},\tau}\right)
\right]  ^{2}%
\end{array}
\right.  . \label{mu}%
\end{equation}
These notations are divided in a part that depends explicitly on the phase
field $\delta \theta_{\mathbf{x},\tau}$ (indicated by a superscript $\left(
\theta \right)  $), and a part that does not. When $\delta \theta_{\mathbf{x}%
,\tau}$ is set to zero in (\ref{Partition Function (2)}), the saddle-point
result is obtained \cite{Devreese Klimin Tempere}. In (\ref{zeta}) and
(\ref{mu}) the two competing effects of the FFLO state are visible. The first
effect is that the imbalance chemical potential $\zeta$ can be lowered by the
term $\mathbf{k.Q}$, which depends on the FFLO momentum $\mathbf{Q}$. This
shows that the FFLO state is able to cope with spin-imbalance. The second
effect is that the momentum $\mathbf{Q}$ increases the total chemical
potential, i.e. the term $Q^{2}/4$ in $\mu_{\mathbf{Q}}$. This shows that
forming Cooper pairs with non-zero momentum costs more energy than forming
Cooper pairs with zero momentum. The FFLO state is a trade-off between these
two competing effects.

Finally, the fermionic Matsubara summation in (\ref{Partition Function (2)})
can be performed. The resulting action is ultraviolet divergent, which is an
artifact of the contact potential (\ref{Contact potential}). This divergence
can be removed by writing the interaction strength $g$ as a function of the
s-wave scattering length $a_{s}$, as in expression (\ref{1/g(as) (2)}). The
integral over momentum $\mathbf{k}$ in this expression cancels out the
divergence of the action in (\ref{Partition Function (2)}). The latter then
becomes:%
\begin{align}
S  &  =-\int_{0}^{\beta}d\tau \int d\mathbf{x}\int \frac{d\mathbf{k}}{\left(
2\pi \right)  ^{3}}\left \{  \frac{1}{\beta}\log \left[  2\cosh \left(
\beta \tilde{E}_{\mathbf{k,Q}}^{\left(  \theta \right)  }\right)  +2\cosh \left(
\beta \tilde{\zeta}_{\mathbf{k,Q}}^{\left(  \theta \right)  }\right)  \right]
-\tilde{\xi}_{\mathbf{k,Q}}^{\left(  \theta \right)  }-\frac{\Delta^{2}}%
{2k^{2}}\right \} \label{Action (2)}\\
&  -\beta V\frac{\Delta^{2}}{8\pi a_{s}},\nonumber
\end{align}
with $\tilde{E}_{\mathbf{k,Q}}^{\left(  \theta \right)  }=\sqrt{\left(
\tilde{\xi}_{\mathbf{k,Q}}^{\left(  \theta \right)  }\right)  ^{2}+\Delta^{2}}%
$. This is the effective action for the FFLO state in a 3D Fermi gas with spin-imbalance.

\section{The fluctuation action \label{Fluctuation action}}

In the partition function (\ref{Partition Function (2)}), only the path
integral over the fluctuation field $\delta \theta_{\mathbf{x},\tau}$ remains.
However, at this point, the action is still a complicated functional of this
field, for which the path integral cannot be calculated analytically.
Therefore, the action will be expanded up to quadratic order in $\delta
\theta_{\mathbf{x},\tau}$ and its derivatives. At this point we have made two
approximations regarding the fluctuation field: (1) the fluctuations are
assumed to be small compared to the saddle-point value, which justifies an
expansion up to quadratic order in the fluctuation field and (2) the
fluctuations are assumed to be slowly varying in time and space, compared to
the fermionic fields, which forms the basis for the path-integral adiabatic
approximation. The latter approximation implies that the momentum $q$,
associated with the fluctuation field, is also small. For this reason, we will
take into account terms only up to order $q^{4}$ in the action when Fourier
transforming the fluctuation field, which will be done at a later stage in
this work.

By expanding the action (\ref{Action (2)}) up to quadratic order in
$\delta \theta_{\mathbf{x},\tau}$ and its derivatives, the action can be
written as the sum of two parts: $S=S_{sp}+S_{fl}$. The zeroth order term
yields the saddle-point action $S\left(  \delta \theta_{\mathbf{x},\tau
}=0\right)  =S_{sp}$ and the second order term leads to the fluctuation action
$S_{fl}$. The saddle-point action corresponds to the result obtained in
\cite{Devreese Klimin Tempere}. The fluctuation action is given by%
\begin{align}
S_{fl}  &  =-\int_{0}^{\beta}d\tau \int d\mathbf{x}\int \frac{d\mathbf{k}%
}{\left(  2\pi \right)  ^{3}}\left \{  \left(  1-\frac{\xi_{\mathbf{k,Q}}%
}{E_{\mathbf{k,Q}}}X\left(  E_{\mathbf{k,Q}}\right)  \right)  \left(
-\dfrac{1}{4}\left[  \nabla_{\mathbf{x}}\left(  \delta \theta_{\mathbf{x},\tau
}\right)  \right]  ^{2}\right)  \right. \nonumber \\
&  +\frac{1}{2}Y\left(  E_{\mathbf{k,Q}}\right)  \left(  \left[
\nabla_{\mathbf{x}}\left(  \delta \theta_{\mathbf{x},\tau}\right)
.\mathbf{k}\right]  ^{2}-i\nabla_{\mathbf{x}}^{2}\left(  \delta \theta
_{\mathbf{x},\tau}\right)  \nabla_{\mathbf{x}}\left(  \delta \theta
_{\mathbf{x},\tau}\right)  .\mathbf{k-}\frac{1}{4}\left[  \nabla_{\mathbf{x}%
}^{2}\left(  \delta \theta_{\mathbf{x},\tau}\right)  \right]  ^{2}\right)
\nonumber \\
&  +\frac{1}{2}\left[  \left(  \dfrac{\xi_{\mathbf{k,Q}}}{E_{\mathbf{k,Q}}%
}\right)  ^{2}Y\left(  E_{\mathbf{k,Q}}\right)  +X\left(  E_{\mathbf{k,Q}%
}\right)  \dfrac{\Delta^{2}}{E_{\mathbf{k,Q}}^{3}}\right] \nonumber \\
&  \times \left[  \left(  \nabla_{\mathbf{x}}\left(  \delta \theta
_{\mathbf{x},\tau}\right)  .\dfrac{\mathbf{Q}}{2}\right)  ^{2}+\dfrac{i}%
{2}\nabla_{\mathbf{x}}\left(  \delta \theta_{\mathbf{x},\tau}\right)
.\mathbf{Q}\dfrac{\partial \delta \theta_{\mathbf{x},\tau}}{\partial \tau}%
-\frac{1}{4}\left(  \dfrac{\partial \delta \theta_{\mathbf{x},\tau}}%
{\partial \tau}\right)  ^{2}\right] \nonumber \\
&  +\tilde{Y}\left(  E_{\mathbf{k,Q}}\right)  \left(  \dfrac{\xi
_{\mathbf{k,Q}}}{E_{\mathbf{k,Q}}}\right)  \left[  \left(  \nabla_{\mathbf{x}%
}\left(  \delta \theta_{\mathbf{x},\tau}\right)  .\dfrac{\mathbf{Q}}{2}\right)
\left[  \nabla_{\mathbf{x}}\left(  \delta \theta_{\mathbf{x},\tau}\right)
.\mathbf{k}\right]  -\dfrac{i}{4}\nabla_{\mathbf{x}}\left(  \delta
\theta_{\mathbf{x},\tau}\right)  .\mathbf{Q~}\nabla_{\mathbf{x}}^{2}\left(
\delta \theta_{\mathbf{x},\tau}\right)  \right. \nonumber \\
&  \left.  \left.  +\dfrac{i}{2}\dfrac{\partial \delta \theta_{\mathbf{x},\tau}%
}{\partial \tau}\nabla_{\mathbf{x}}\left(  \delta \theta_{\mathbf{x},\tau
}\right)  .\mathbf{k}\right]  \right \}  , \label{Fluctuation action (2)}%
\end{align}
where the following notations were used:%
\begin{equation}
\left \{
\begin{array}
[c]{l}%
\smallskip X\left(  A\right)  =\dfrac{\sinh \left(  \beta A\right)  }%
{\cosh \left(  \beta E_{\mathbf{k,Q}}\right)  +\cosh \left(  \beta
\zeta_{\mathbf{k,Q}}\right)  }\\
\smallskip Y\left(  E_{\mathbf{k,Q}}\right)  =\beta \dfrac{1+\cosh \left(  \beta
E_{\mathbf{k,Q}}\right)  \cosh \left(  \beta \zeta_{\mathbf{k,Q}}\right)
}{\left[  \cosh \left(  \beta E_{\mathbf{k,Q}}\right)  +\cosh \left(  \beta
\zeta_{\mathbf{k,Q}}\right)  \right]  ^{2}}\\
\smallskip \tilde{Y}\left(  E_{\mathbf{k,Q}}\right)  =\dfrac{\beta \sinh \left(
\beta E_{\mathbf{k,Q}}\right)  \sinh \left(  \beta \zeta_{\mathbf{k,Q}}\right)
}{\left[  \cosh \left(  \beta E_{\mathbf{k,Q}}\right)  +\cosh \left(  \beta
\zeta_{\mathbf{k,Q}}\right)  \right]  ^{2}}%
\end{array}
\right.  . \label{X,Y,Y_tilde}%
\end{equation}
More details about the derivation of (\ref{Fluctuation action (2)}) are given
in appendix \ref{Obtaining fluctuation action}. As a last step before
calculating the path integral over the fluctuation field $\delta
\theta_{\mathbf{x},\tau}$, we have to perform a Fourier transformation of this
field, again to eliminate the derivatives in the action
(\ref{Fluctuation action (2)}). The result is:%
\begin{align}
S_{fl}  &  =\frac{1}{2}\sum_{\mathbf{q},m}\left \{  \int \frac{d\mathbf{k}%
}{\left(  2\pi \right)  ^{3}}\left[  \dfrac{1}{2}\left(  1-\frac{\xi
_{\mathbf{k,Q}}}{E_{\mathbf{k,Q}}}X\left(  E_{\mathbf{k,Q}}\right)  \right)
q^{2}-Y\left(  E_{\mathbf{k,Q}}\right)  \left(  \left(  \mathbf{q.\mathbf{k}%
}\right)  ^{2}\mathbf{-}\frac{1}{4}q^{4}\right)  \right.  \right. \nonumber \\
&  -\left(  \left(  \dfrac{\xi_{\mathbf{k,Q}}}{E_{\mathbf{k,Q}}}\right)
^{2}Y\left(  E_{\mathbf{k,Q}}\right)  +X\left(  E_{\mathbf{k,Q}}\right)
\dfrac{\Delta^{2}}{E_{\mathbf{k,Q}}^{3}}\right)  \left(  \frac{1}{4}\left(
\mathbf{q.Q}\right)  ^{2}-\frac{1}{2}i\varpi_{m}\left(  \mathbf{q.Q}\right)
-\frac{1}{4}\varpi_{m}^{2}\right) \nonumber \\
&  \left.  \left.  -\tilde{Y}\left(  E_{\mathbf{k,Q}}\right)  \left[  \left(
\mathbf{q.Q}\right)  \left(  \mathbf{q.\mathbf{k}}\right)  -i\varpi_{m}\left(
\mathbf{q.k}\right)  \right]  \medskip \right]  \medskip \right \}  \delta
\theta_{\mathbf{q},m}\delta \theta_{\mathbf{q},m}^{\ast}.
\label{Action - fluctuation (2)}%
\end{align}
In expression (\ref{Action - fluctuation (2)}), $\mathbf{q}$ is the momentum
of the fluctuation field and $\varpi_{m}=2\pi m/\beta$ are bosonic Matsubara
frequencies. The action in (\ref{Action - fluctuation (2)}) contains three
scalar products: $\mathbf{k.q}$, $\mathbf{q.Q}$ and $\mathbf{k.Q}$. When
calculating the integrals over momentum $\mathbf{k}$ and $\mathbf{q}$, one has
to be careful when integrating over the polar and azimuthal angles in
spherical coordinates. Here, for the benefit of clarity, we explicitly mention
our conventions in the definition of these various angles. Within the
integration over $\mathbf{k}$, $\mathbf{Q}$ points along the z-axis, such that
$\mathbf{k.Q=}\left \vert \mathbf{k}\right \vert \left \vert \mathbf{Q}%
\right \vert $cos$\left(  \alpha_{\mathbf{k,Q}}\right)  $ with $\alpha
_{\mathbf{k,Q}}$ the polar angle of the spherical coordinate system.
Analogously, the scalar product of $\mathbf{q}$ and $\mathbf{Q}$ is written
as: $\mathbf{q.Q=}\left \vert \mathbf{q}\right \vert \left \vert \mathbf{Q}%
\right \vert \cos \left(  \alpha_{\mathbf{q,Q}}\right)  $. The vector
$\mathbf{q}$, however, lies in an arbitrary direction with respect to the
vector $\mathbf{k}$. As a consequence, the scalar product of these two vectors
is given by $\mathbf{k.q=}\left \vert \mathbf{k}\right \vert \left \vert
\mathbf{q}\right \vert \cos \left(  \alpha_{\mathbf{k,q}}\right)  $, where the
angle $\alpha_{\mathbf{k,q}}$ can be written in terms of both the polar and
the azimuthal angle between $\mathbf{k}$ and $\mathbf{Q}$ on the one hand and
between $\mathbf{q}$ and $\mathbf{Q}$ on the other hand, by using:%
\begin{equation}
\cos \left(  \alpha_{\mathbf{k,q}}\right)  =xy+\sqrt{1-x^{2}}\sqrt{1-y^{2}}%
\cos \left(  \varphi_{\mathbf{k,Q}}-\varphi_{\mathbf{q,Q}}\right)  ,
\label{theta_k,q}%
\end{equation}
where $x=\cos \left(  \alpha_{\mathbf{k,Q}}\right)  $ and $y=\cos \left(
\alpha_{\mathbf{q,Q}}\right)  $. Substituting (\ref{theta_k,q}) in the action
(\ref{Action - fluctuation (2)}) and integrating out $\varphi_{\mathbf{k,Q}}$,
the fluctuation partition function, defined by $\mathcal{Z}_{fl}=\exp \left(
-S_{fl}\right)  $ becomes%
\begin{equation}
\mathcal{Z}_{fl}=\int \mathcal{D}\delta \theta_{\mathbf{q},m}\exp \left(
-\frac{1}{2}\sum_{\mathbf{q},m}\delta \theta_{\mathbf{q},m}^{\ast}%
\mathbb{A}_{\mathbf{q},m}\delta \theta_{\mathbf{q},m}\right)  .
\label{Partition Function (3)}%
\end{equation}
Here, $\mathbb{A}_{\mathbf{q},m}$ equals%
\begin{equation}
\mathbb{A}_{\mathbf{q},m}=A_{\mathbf{Q},x}~q^{2}+B_{\mathbf{Q},x}~q^{2}%
y^{2}+C_{\mathbf{Q},x}~q^{4}+D_{\mathbf{Q},x}~i\varpi_{m}~qy-E_{\mathbf{Q}%
,x}~\left(  i\varpi_{m}\right)  ^{2},
\end{equation}
where the following coefficients have been introduced%
\begin{equation}
\left \{
\begin{array}
[c]{l}%
\smallskip A_{\mathbf{Q},x}=\dfrac{1}{2}%
{\displaystyle \int}
\dfrac{d\mathbf{k}}{\left(  2\pi \right)  ^{3}}\left(  1-\dfrac{\xi
_{\mathbf{k,Q}}}{E_{\mathbf{k,Q}}}X\left(  E_{\mathbf{k,Q}}\right)
-k^{2}\left(  1-x^{2}\right)  Y\left(  E_{\mathbf{k,Q}}\right)  \right) \\
\smallskip B_{\mathbf{Q},x}=%
{\displaystyle \int}
\dfrac{d\mathbf{k}}{\left(  2\pi \right)  ^{3}}\left \{  \dfrac{1}{2}%
k^{2}\left(  1-3x^{2}\right)  Y\left(  E_{\mathbf{k,Q}}\right)  -kQx\tilde
{Y}\left(  E_{\mathbf{k,Q}}\right)  \left(  \dfrac{\xi_{\mathbf{k,Q}}%
}{E_{\mathbf{k,Q}}}\right)  \right. \\
\smallskip \hspace{0.5in}\left.  -\dfrac{Q^{2}}{4}\left[  \left(  \dfrac
{\xi_{\mathbf{k,Q}}}{E_{\mathbf{k,Q}}}\right)  ^{2}Y\left(  E_{\mathbf{k,Q}%
}\right)  +X\left(  E_{\mathbf{k,Q}}\right)  \dfrac{\Delta^{2}}%
{E_{\mathbf{k,Q}}^{3}}\right]  \right \} \\
\smallskip C_{\mathbf{Q},x}=\dfrac{1}{4}%
{\displaystyle \int}
\dfrac{d\mathbf{k}}{\left(  2\pi \right)  ^{3}}Y\left(  E_{\mathbf{k,Q}}\right)
\\
\smallskip D_{\mathbf{Q},x}=%
{\displaystyle \int}
\dfrac{d\mathbf{k}}{\left(  2\pi \right)  ^{3}}\left \{  \dfrac{Q}{2}\left[
\left(  \dfrac{\xi_{\mathbf{k,Q}}}{E_{\mathbf{k,Q}}}\right)  ^{2}Y\left(
E_{\mathbf{k,Q}}\right)  +X\left(  E_{\mathbf{k,Q}}\right)  \dfrac{\Delta^{2}%
}{E_{\mathbf{k,Q}}^{3}}\right]  \right. \\
\smallskip \hspace{0.5in}\left.  +kx\tilde{Y}\left(  E_{\mathbf{k,Q}}\right)
\left(  \dfrac{\xi_{\mathbf{k,Q}}}{E_{\mathbf{k,Q}}}\right)  \right \} \\
\smallskip E_{\mathbf{Q},x}=\dfrac{1}{4}%
{\displaystyle \int}
\dfrac{d\mathbf{k}}{\left(  2\pi \right)  ^{3}}\left[  \left(  \dfrac
{\xi_{\mathbf{k,Q}}}{E_{\mathbf{k,Q}}}\right)  ^{2}Y\left(  E_{\mathbf{k,Q}%
}\right)  +X\left(  E_{\mathbf{k,Q}}\right)  \dfrac{\Delta^{2}}%
{E_{\mathbf{k,Q}}^{3}}\right]
\end{array}
\right.  .
\end{equation}

At this point only the path integral over the fluctuation field remains in the
partition function given by expression (\ref{Partition Function (3)}). When
calculating this path integral, one has to be careful not to double-count the
fields. The reason is that $\delta \theta_{\mathbf{q},m}=\delta \theta
_{-\mathbf{q},-m}^{\ast}$ because $\delta \theta_{\mathbf{x},\tau}$ is a real
function. To circumvent this problem, only half the total momentum domain is
taken into account:
\begin{equation}
\mathcal{Z}_{fl}=\prod_{\substack{\mathbf{q},m\\q_{z}>0}}\int d\delta
\theta_{\mathbf{q},m}\int d\delta \theta_{\mathbf{q},m}^{\ast}\exp \left(
-\sum_{_{\substack{\mathbf{q},m\\q_{z}>0}}}\delta \theta_{\mathbf{q},m}^{\ast
}\mathbb{A}_{\mathbf{q},m}\delta \theta_{\mathbf{q},m}\right)  ,
\end{equation}
where the symmetry property%
\begin{equation}
\mathbb{A}_{\mathbf{q},m}=\mathbb{A}_{-\mathbf{q},-m},
\end{equation}
was used. Now, the standard expression for a quadratic bosonic path integral
can be used, which leads to
\begin{equation}
\mathcal{Z}_{fl}=\exp \left(  -\frac{1}{2}\sum_{\mathbf{q},m}\ln \left[
-E_{\mathbf{Q},x}~\left(  i\varpi_{m}\right)  ^{2}+D_{\mathbf{Q},x}%
~i\varpi_{m}~qy+A_{\mathbf{Q},x}~q^{2}+B_{\mathbf{Q},x}~q^{2}y^{2}%
+C_{\mathbf{Q},x}~q^{4}\right]  \right)  .
\end{equation}
As a final step, the bosonic Matsubara summation can be calculated, which
eventually results in the fluctuation free energy:%
\begin{equation}
\Omega_{fl}\left(  \mu,\zeta,\beta;\Delta,Q\right)  =\frac{1}{2\beta V}%
\sum_{\mathbf{q}}\left \{  \ln \left[  2\cosh \left(  \beta \sqrt{\Psi}\right)
-2\cosh \left(  \beta \Xi \right)  \right]  -\beta \sqrt{\Psi}\right \}  ,
\label{Free energy}%
\end{equation}
with%
\begin{equation}
\left \{
\begin{array}
[c]{l}%
\sqrt{\Psi}=\dfrac{1}{2E_{\mathbf{Q},x}}\sqrt{D_{\mathbf{Q},x}^{2}q^{2}%
y^{2}+4E_{\mathbf{Q},x}\left(  A_{\mathbf{Q},x}~q^{2}+B_{\mathbf{Q},x}%
~q^{2}y^{2}+C_{\mathbf{Q},x}~q^{4}\right)  }\\
\Xi=\dfrac{D_{\mathbf{Q},x}}{2E_{\mathbf{Q},x}}qy
\end{array}
\right.  .
\end{equation}

\section{The T%
$>$%
0 phase diagram \label{Phase diagram}}

\subsection{Method and caveat}

In this subsection we briefly explain the method for constructing the phase
diagram of the system, including phase fluctuations, starting from the
fluctuation free energy (\ref{Free energy}). In our previous work \cite{Review
FFLO 3D Devreese et al}, the saddle-point phase diagram for a 3D
spin-imbalanced Fermi gas at zero temperature was calculated, as a function of
the chemical potentials $\mu$ and $\zeta$. For a given value of $\mu$ and
$\zeta$, the saddle-point free energy was minimized with respect to the
variational parameters $\Delta$ (the superfluid band gap) and $Q$ (the FFLO
momentum). The values of $\Delta$ and $Q$ at a given minimum then determined
which state this minimum corresponds to: the BCS state ($\Delta \neq0,Q=0$),
the FFLO state ($\Delta \neq0,Q\neq0$) or the normal state ($\Delta=0$). Now,
phase fluctuations around the saddle point can be included by using the
fluctuation free energy (\ref{Free energy}). For each value of $\mu$ and
$\zeta$, the contribution of the fluctuation free energy to the different
minima of the saddle-point free energy can be calculated. This will result in
a shift of these minima relative to each other, leading to corrections to the
phase diagram.

However, an important caveat has to be kept in mind. In this paper, phase
fluctuations are calculated by using an expansion of the action around the
saddle point up to quadratic order in the fluctuation field. This expansion
only makes sense when the saddle point is a local or a global minimum. If this
is not the case, the quadratic terms will result in negative contributions to
the action, which leads to an exponential divergence of the partition function
because $\mathcal{Z}\sim e^{-S}$. Because of this fact, only corrections to
the BCS-FFLO transition and to the BCS-normal transition can be calculated
within the current formalism, whereas this is not possible for the FFLO-normal
transition. The reason for this is that the former two transitions are both of
first order, which means there is a competition between two local minima. For
given values of $\mu$ and $\zeta$, the fluctuation corrections to both these
minima can be calculated and thus their relative shift can be determined. In
contrast, the FFLO-normal transition is of second order, so that there is a
continuous transition between the FFLO minimum and the normal minimum. From
this it follows that for given values of $\mu$ and $\zeta$, either the FFLO
minimum or the normal minimum is present, but never both at the same time.
Hence, fluctuation corrections can only be calculated for one of the two
states, for given values of $\mu$ and $\zeta$. As a result, it is not possible
to calculate corrections to the FFLO-normal transition.

Finally it should be mentioned that in order to calculate fluctuation
corrections to the normal state, both amplitude- and phase fluctuations should
be taken into account. This is because when $\Delta$ goes to zero, both types
of fluctuations become indistinguishable. Because of this reason, we will
focus solely on fluctuation corrections to the BCS-FFLO transition in this paper.

\subsection{The $\left(  \mu,\zeta \right)  $-phase diagram at $T>0$}

\begin{figure}
[h]
\begin{center}
\includegraphics[
height=8.3933cm,
width=12.7931cm
]%
{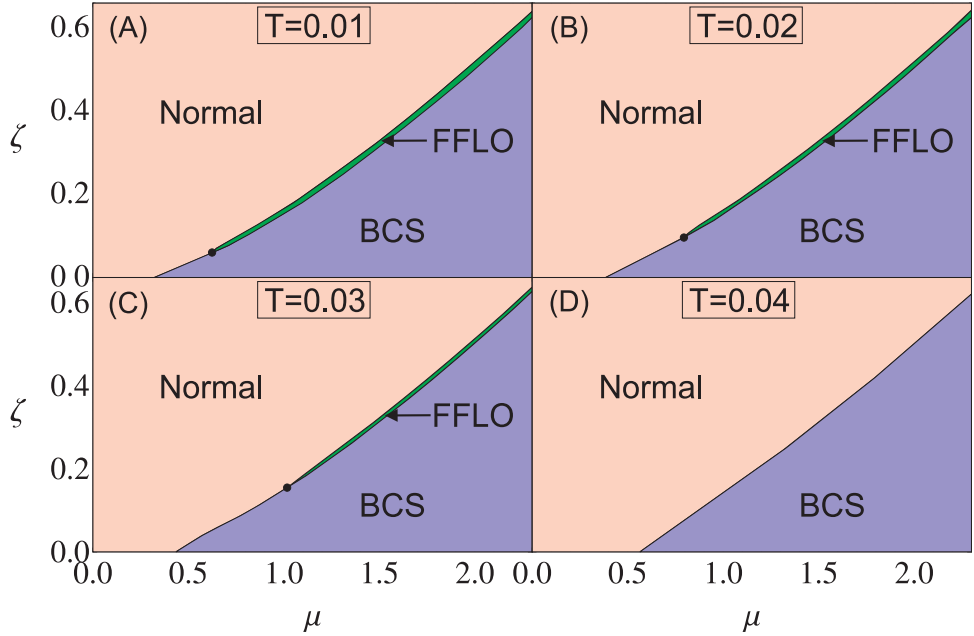}%
\caption{The saddle-point phase diagram of a 3D Fermi gas with spin-imbalance,
as a function of the total chemical potential $\mu$ and the imbalance chemical
potential $\zeta$, for several values of temperature $T$ (in units
$\hbar=2m=\left \vert a_{s}\right \vert =1$). The tri-critical point at which
the BCS-normal transition meets the BCS-FFLO transition is indicated by a
black dot in figures (A), (B) and (C). The presence of the FFLO state
decreases when temperature increases, and at $T\approx0.04$, the FFLO state
vanishes for all values of $\mu$. It should be noted that $T/T_{F}$ is not
constant in these phase diagrams, because we use $\left \vert a_{s}\right \vert
$ as a unit of length instead of $k_{F}$. When converting temperature to units
of the Fermi temperature, we find that the tri-critical point lies at
$T/T_{F}\approx0.03$.}%
\label{fdmuzeta_fl_3d_english.eps}%
\end{center}
\end{figure}

Figure \ref{fdmuzeta_fl_3d_english.eps} shows the saddle-point phase diagram
of a 3D Fermi gas with spin-imbalance, as a function of the chemical
potentials $\mu$ and $\zeta$, at different temperatures.%
Three different states are indicated: the BCS state (light blue area), the
FFLO state (light green area), and the normal state (light red area). As is
well known, at the saddle-point level, the presence of the FFLO state in the
phase diagram of a spin-imbalanced 3D Fermi gas is minimal. With increasing
temperature, the area of the FFLO state decreases in size, because the free
energy of the normal state lowers relatively to the free energy of the FFLO
state (as well as the BCS state) when temperature increases. We find that at
$T=0.04$, the FFLO state vanishes for all values of $\mu$.

\begin{figure}
[h]
\begin{center}
\includegraphics[
trim=0.702415in -0.000007in -0.702414in 0.000007in,
height=7.6047cm,
width=12.7931cm
]%
{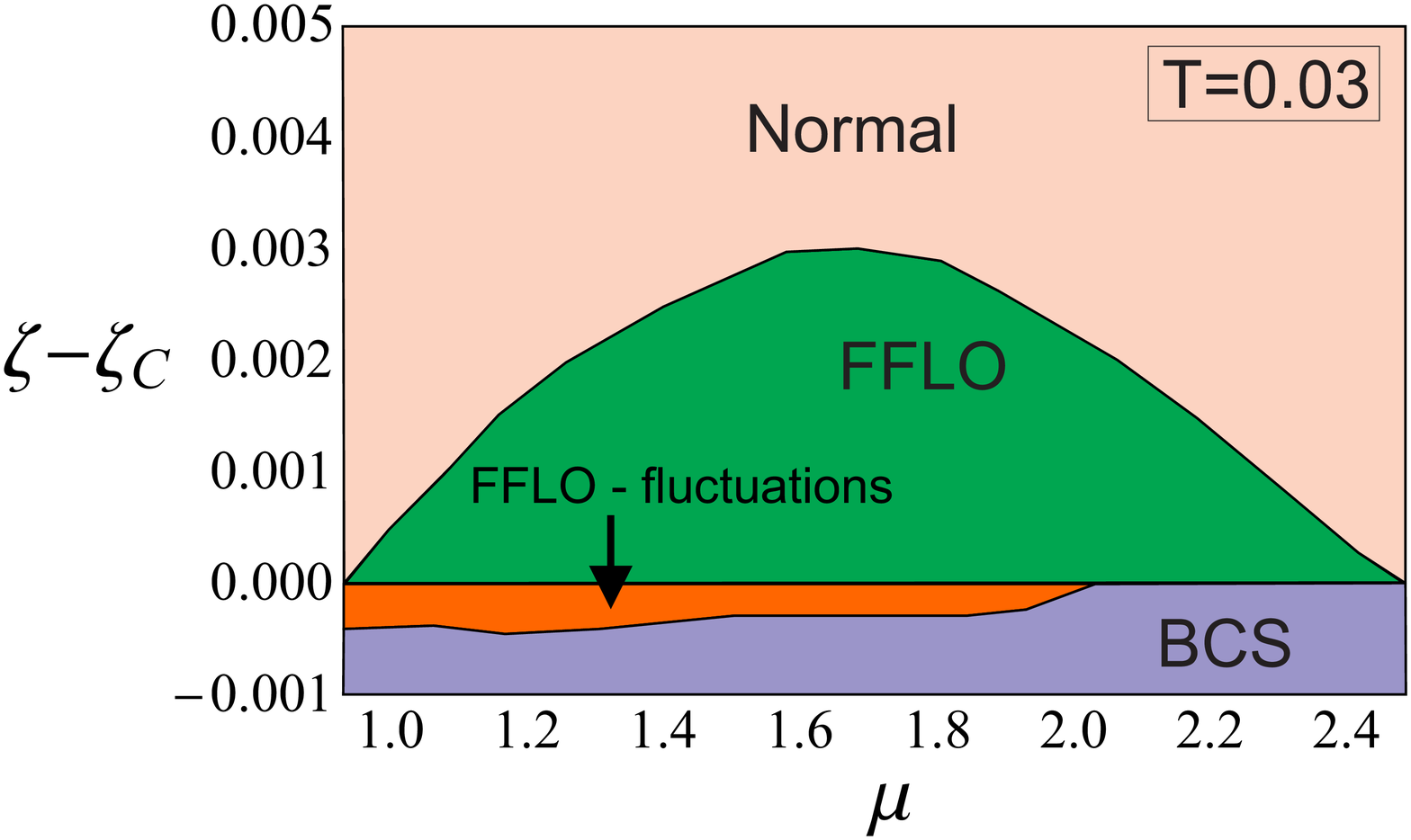}%
\caption{Phase diagram of a 3D Fermi gas with spin-imbalance at $T=0.03$ (in
units $\hbar=2m=\left \vert a_{s}\right \vert =1$). This phase diagram is a
rescaled version of fig. \ref{fdmuzeta_fl_3d_english.eps}(C), with the
addition of corrections due to phase fluctuations. The rescaling is performed
by plotting $\zeta$ relative to the value of $\zeta$ at which the BCS-FFLO
transition occurs ($\zeta_{C}$) at the saddle-point level. The fluctuation
corrections lead to a small enlargement of the FFLO region (green + orange
area) compared to the saddle-point result (green area). Given the size of the
FFLO area in fig. \ref{fdmuzeta_fl_3d_english.eps}(c), this figure
demonstrates that fluctuations only lead to small quantitative corrections to
the FFLO area. We did not plot lower values than $\mu=1.0$ because we only
calculated corrections to the BCS-FFLO transition. At higher values than
$\mu=2.0$, the contribution of fluctuations is negligible, even at the scale
of this figure. The somewhat irregular shape of the fluctuation area is due to
the resolution of our numerical grid.}%
\label{fdmuzeta_fl_t003_details.eps}%
\end{center}
\end{figure}

When adding fluctuation corrections to these phase diagrams, we find,
surprisingly, that these corrections are very small so that their influence is
not visible at the scale of fig. \ref{fdmuzeta_fl_3d_english.eps}. To give an
idea of the order of magnitude of the fluctuation corrections, we `zoom in' on
the FFLO area at $T=0.03$ by re-scaling the ordinate of the phase diagram in
fig. \ref{fdmuzeta_fl_3d_english.eps}(C). More specifically, we use
$\zeta-\zeta_{C}$ instead of $\zeta$, where $\zeta_{C}=\zeta_{C}\left(
\mu \right)  $ is the value of $\zeta$ (for a given value of $\mu$) at which
the BCS-FFLO transition occurs within the saddle-point approximation. Using
this rescaling method, the phase diagram at $T=0.03$, now including
fluctuation corrections, is given by fig. \ref{fdmuzeta_fl_t003_details.eps}.
We find that the FFLO area is slightly enlarged when fluctuations are taken
into account (green + orange area) compared to the saddle-point result (green
area), i.e. the BCS-FFLO transition line lies at lower values of $\zeta
-\zeta_{C}$. The orange (fluctuation) area is relatively small, even compared
to the FFLO area, which in itself is only a small sliver on the phase diagram,
as can be seen in fig. \ref{fdmuzeta_fl_3d_english.eps}. Thus we find that
phase fluctuations, at least within a hydrodynamic approach, have only a
relatively small quantitative effect on the FFLO state, when considering the
BCS-FFLO transition. In fig. \ref{fdmuzeta_fl_t003_details.eps} we have
omitted the values of $\mu$ at which the FFLO state is not present, since we
only calculated corrections for the BCS-FFLO transition and not for the
BCS-normal transition. The somewhat irregular shape of the fluctuation area is
due to the resolution of our numerical grid. At higher values than $\mu=2.0$,
the contribution of fluctuations is negligible, even at the scale of this
figure. This may seem counterintuitive, because the interaction strength
increases when the density and hence the total chemical potential $\mu$ is
increased, and fluctuations are expected to contribute more at higher values
of the interaction strength. The reason, however, is that because we use the
scattering length as a unit of length (instead of the Fermi vector $k_{F}$)
the temperature relative to the Fermi temperature $T/T_{F}$ is not a constant
in figure \ref{fdmuzeta_fl_t003_details.eps}: $T/T_{F}$ decreases with
increasing $\mu$, because the density and hence $T_{F}=k_{F}^{2}/k_{B}$ (in
units $\hbar=2m=1$) increases when $\mu$ is increased. This explains why the
fluctuation corrections, which decrease in magnitude at lower temperatures,
become smaller for increasing values of $\mu$ in fig.
\ref{fdmuzeta_fl_t003_details.eps}(C). When converting the temperature in fig.
\ref{fdmuzeta_fl_3d_english.eps} to\ units of the Fermi temperature, we find
that the tri-critical point (indicated by a black dot in this figure) lies at
approximately $T/T_{F}\approx0.03$.

\section{Conclusions \label{Conclusions}}

In this paper, we have studied the effect of phase fluctuations on the FFLO
state in a 3D Fermi gas with spin-imbalance. Starting from the partition
function of the system, the complex fields of the Cooper pairs were written in
terms of an amplitude field and a phase field. By choosing a suitable saddle
point, in which Cooper pairs were allowed to have a non-zero momentum
$\mathbf{Q}$, the FFLO state was included into the mathematical description.
We then considered fluctuations of the phase field around the saddle point.
Because the resulting path integrals over the fermionic fields and the
fluctuation field could not be calculated exactly, we used the path-integral
adiabatic approximation. Subsequently, the action was expanded up to second
order in the fluctuation field, after which the path integral over this field
could be calculated. This led to the fluctuation free energy. This free energy
was used to calculate corrections to the saddle-point phase diagram of the
system. For each value of the chemical potentials $\mu$ and $\zeta$, the
contribution of the fluctuation free energy to each minimum of the
saddle-point free energy was calculated. This method allows to calculate
corrections to the BCS-FFLO transition, but not to the FFLO-normal transition.
We have found that phase fluctuations only lead to relatively small
quantitative corrections to the BCS-FFLO transition. Our results suggest that
the fluctuation of the phase of the FFLO order parameter, which can be
interpreted as an oscillation of the FFLO momentum vector around its
saddle-point value, does not cause an instability of the FFLO state with
respect to the BCS state.

\begin{description}
\item[Acknowledgements] We wish to thank Carlos S\'{a} de Melo, Nick
Proukakis, Michiel Wouters, Fons Brosens and Serghei Klimin for interesting
and stimulating discussions. JPAD gratefully acknowledges a Post-doctoral
fellowship of the Research Foundation - Flanders (FWO-Vlaanderen). This work
was supported by FWO-Vlaanderen projects G.0119.12.N, G.0115.12.N, and G.0180.09.N.
\end{description}

\appendix{}

\section{Fourier transformation within the path-integral adiabatic approach
\label{Fourier transform}}

When applying the averaging given by (\ref{PIAA}) to the partition function
(\ref{Partitiesom FL}), the latter becomes%
\begin{align}
\mathcal{Z}  &  \mathcal{=}\prod_{\mathbf{x},\tau,\sigma}\prod_{\mathbf{x}%
^{\prime},\tau^{\prime}}\int d\bar{\psi}_{\mathbf{x}^{\prime},\tau^{\prime
},\sigma}^{\left(  \mathbf{x},\tau \right)  }d\psi_{\mathbf{x}^{\prime}%
,\tau^{\prime},\sigma}^{\left(  \mathbf{x},\tau \right)  }\int \mathcal{D}%
\delta \theta_{\mathbf{x},\tau}\nonumber \\
&  \times \exp \left(  -\int_{0}^{\beta}d\tau \int d\mathbf{x}\frac{1}{\beta
V}\mathbf{~}\int_{0}^{\beta}d\tau^{\prime}\int d\mathbf{x}^{\prime}~\bar{\eta
}_{\mathbf{x}^{\prime},\tau^{\prime}}^{\left(  \mathbf{x},\tau \right)
}\left \{  -\mathbb{G}_{\mathbf{x}^{\prime},\tau^{\prime}}^{-1}\left[
\delta \theta \left(  \mathbf{x},\tau \right)  \right]  \right \}  \eta
_{\mathbf{x}^{\prime},\tau^{\prime}}^{\left(  \mathbf{x},\tau \right)  }+\beta
V\frac{\Delta^{2}}{g}\right)  , \label{A - Partition function (2)}%
\end{align}
where the Nambu spinors are given by (\ref{Nambu spinor}). Let us first
transform the diagonal terms of the inverse Green's function, given by
expression (\ref{-G^-1 x,t (2)}). Our definitions for the Fourier transform
are as follows:%
\begin{equation}
\left \{
\begin{array}
[c]{c}%
\bar{\psi}_{\mathbf{x}^{\prime},\tau^{\prime},\sigma}^{\left(  \mathbf{x}%
,\tau \right)  }=\left(  \beta V\right)  ^{-\frac{1}{2}}\sum_{\mathbf{k}%
,\omega_{n}}e^{i\omega_{n}\tau^{\prime}-i\mathbf{k.x}^{\prime}}\bar{\psi
}_{\mathbf{k},\omega_{n},\sigma}^{\left(  \mathbf{x},\tau \right)  }\\
\psi_{\mathbf{x}^{\prime},\tau^{\prime},\sigma}^{\left(  \mathbf{x}%
,\tau \right)  }=\left(  \beta V\right)  ^{-\frac{1}{2}}\sum_{\mathbf{k}%
,\omega_{n}}e^{-i\omega_{n}\tau^{\prime}+i\mathbf{k.x}^{\prime}}%
\psi_{\mathbf{k},\omega_{n},\sigma}^{\left(  \mathbf{x},\tau \right)  }%
\end{array}
\right.  .
\end{equation}
The term proportional to $\bar{\psi}_{\uparrow}\psi_{\uparrow}$ can be
transformed as
\begin{align}
&  -\int_{0}^{\beta}d\tau \int d\mathbf{x}\frac{1}{\beta V}\int_{0}^{\beta
}d\tau^{\prime}\int d\mathbf{x}~\bar{\psi}_{\mathbf{x}^{\prime},\tau^{\prime
},\uparrow}^{\left(  \mathbf{x},\tau \right)  }\left(
\begin{array}
[c]{c}%
\dfrac{\partial}{\partial \tau}-\nabla_{\mathbf{x}}^{2}-\mu+\dfrac{Q^{2}}%
{4}-\zeta-i\mathbf{Q}.\nabla_{\mathbf{x}}%
\end{array}
\right. \nonumber \\
&  \left.  +\dfrac{i}{2}\dfrac{\partial \delta \theta_{\mathbf{x},\tau}%
}{\partial \tau}-i\nabla_{\mathbf{x}}\left(  \delta \theta_{\mathbf{x},\tau
}\right)  \nabla_{\mathbf{x}}-\dfrac{i}{2}\nabla_{\mathbf{x}}^{2}\left(
\delta \theta_{\mathbf{x},\tau}\right)  +\nabla_{\mathbf{x}}\left(
\delta \theta_{\mathbf{x},\tau}\right)  .\dfrac{\mathbf{Q}}{2}+\dfrac{1}%
{4}\left[  \nabla_{\mathbf{x}}\left(  \delta \theta_{\mathbf{x},\tau}\right)
\right]  ^{2}\right)  \psi_{\mathbf{x}^{\prime},\tau^{\prime},\uparrow
}^{\left(  \mathbf{x},\tau \right)  }\nonumber \\
&  =-\frac{1}{\beta V}\int_{0}^{\beta}d\tau \int d\mathbf{x}\sum_{\mathbf{k}%
,\omega_{n}}\left(  -i\omega_{n}+k^{2}-\mu+\dfrac{Q^{2}}{4}-\zeta
+\mathbf{Q}.\mathbf{k}\right. \nonumber \\
&  \left.  +\dfrac{i}{2}\dfrac{\partial \delta \theta_{\mathbf{x},\tau}%
}{\partial \tau}+\nabla_{\mathbf{x}}\left(  \delta \theta_{\mathbf{x},\tau
}\right)  .\mathbf{k}-\dfrac{i}{2}\nabla_{\mathbf{x}}^{2}\left(  \delta
\theta_{\mathbf{x},\tau}\right)  +\nabla_{\mathbf{x}}\left(  \delta
\theta_{\mathbf{x},\tau}\right)  .\dfrac{\mathbf{Q}}{2}+\dfrac{1}{4}\left[
\nabla_{\mathbf{x}}\left(  \delta \theta_{\mathbf{x},\tau}\right)  \right]
^{2}\right)  \bar{\psi}_{\mathbf{k},\omega_{n},\uparrow}^{\left(
\mathbf{x},\tau \right)  }\psi_{\mathbf{k},\omega_{n},\uparrow}^{\left(
\mathbf{x},\tau \right)  },
\end{align}
and analogously for the term proportional to $\bar{\psi}_{\downarrow}%
\psi_{\downarrow}$. The off-diagonal term becomes%
\begin{equation}
-\int_{0}^{\beta}d\tau \int d\mathbf{x}\frac{1}{\beta V}\int_{0}^{\beta}%
d\tau^{\prime}\int d\mathbf{x}^{\prime}~\bar{\psi}_{\mathbf{x}^{\prime}%
,\tau^{\prime},\uparrow}^{\left(  \mathbf{x},\tau \right)  }\bar{\psi
}_{\mathbf{x}^{\prime},\tau^{\prime},\downarrow}^{\left(  \mathbf{x}%
,\tau \right)  }\Delta=-\int_{0}^{\beta}d\tau \int d\mathbf{x}\frac{1}{\beta
V}\sum_{\mathbf{k},\omega_{n}}\bar{\psi}_{\mathbf{k},\omega_{n},\uparrow
}^{\left(  \mathbf{x},\tau \right)  }\bar{\psi}_{-\mathbf{k},-\omega
_{n},\downarrow}^{\left(  \mathbf{x},\tau \right)  }\Delta.
\end{equation}
Now the partition function has become%
\begin{align}
\mathcal{Z}  &  =\prod_{\mathbf{x},\tau}\prod_{\mathbf{k},\omega_{n},\sigma
}\int d\bar{\psi}_{\mathbf{k},\omega_{n},\sigma}^{\left(  \mathbf{x}%
,\tau \right)  }d\psi_{\mathbf{k},\omega_{n},\sigma}^{\left(  \mathbf{x}%
,\tau \right)  }\prod_{\mathbf{x},\tau}\int \mathcal{D\delta}\theta
_{\mathbf{x},\tau}\nonumber \\
&  \times \exp \left(  -\frac{1}{\beta V}\int_{0}^{\beta}d\tau \int
d\mathbf{x}\sum_{\mathbf{k},\omega_{n}}\mathbf{~}\bar{\eta}_{\mathbf{k}%
,\omega_{n}}^{\left(  \mathbf{x},\tau \right)  }\left \{  -\mathbb{G}%
_{\mathbf{k},\omega_{n}}^{-1}\left[  \delta \theta \left(  \mathbf{x}%
,\tau \right)  \right]  \right \}  \eta_{\mathbf{k},\omega_{n}}^{\left(
\mathbf{x},\tau \right)  }+\beta V\frac{\Delta^{2}}{g}\right)  ,
\label{A -Partition function}%
\end{align}
where $-\mathbb{G}_{\mathbf{k},\omega_{n}}^{-1}\left[  \delta \theta \left(
\mathbf{x},\tau \right)  \right]  $ is given by%
\begin{align}
-\mathbb{G}_{\mathbf{k},\omega_{n}}^{-1}\left[  \delta \theta \left(
\mathbf{x},\tau \right)  \right]   &  =\left(  -i\omega_{n}-\zeta
+\mathbf{Q}.\mathbf{k}+\nabla_{\mathbf{x}}\left(  \delta \theta_{\mathbf{x}%
,\tau}\right)  .\mathbf{k}-\dfrac{i}{2}\nabla_{\mathbf{x}}^{2}\left(
\delta \theta_{\mathbf{x},\tau}\right)  \right)  \sigma_{0}\nonumber \\
&  +\left(  k^{2}-\mu+\dfrac{Q^{2}}{4}+\dfrac{i}{2}\dfrac{\partial \delta
\theta_{\mathbf{x},\tau}}{\partial \tau}+\nabla_{\mathbf{x}}\left(
\delta \theta_{\mathbf{x},\tau}\right)  .\dfrac{\mathbf{Q}}{2}+\dfrac{1}%
{4}\left[  \nabla_{\mathbf{x}}\left(  \delta \theta_{\mathbf{x},\tau}\right)
\right]  ^{2}\right)  \sigma_{3}+\Delta \sigma_{1},
\end{align}
with the Nambu-spinors%
\begin{equation}
\bar{\eta}_{\mathbf{k},\omega_{n}}^{\left(  \mathbf{x},\tau \right)  }=\left(
\begin{array}
[c]{cc}%
\bar{\psi}_{\mathbf{k},\omega_{n},\uparrow}^{\left(  \mathbf{x},\tau \right)  }
& \psi_{-\mathbf{k},-\omega_{n},\downarrow}^{\left(  \mathbf{x},\tau \right)  }%
\end{array}
\right)  \text{ en }\eta_{\mathbf{k}^{\prime},\omega_{n}^{\prime}}^{\left(
\mathbf{x},\tau \right)  }=\left(
\begin{array}
[c]{c}%
\psi_{\mathbf{k},\omega_{n},\uparrow}^{\left(  \mathbf{x},\tau \right)  }\\
\bar{\psi}_{-\mathbf{k},-\omega_{n},\downarrow}^{\left(  \mathbf{x}%
,\tau \right)  }%
\end{array}
\right)  .
\end{equation}
Note that in (\ref{A -Partition function}) the term proportional to $\bar
{\psi}_{\downarrow}\psi_{\downarrow}$ has had its sum over momentum re-indexed
as follows: $\mathbf{k\rightarrow-k}$.

\section{Obtaining the fluctuation action after Fourier expansion of the
effective hydrodynamic action \label{Obtaining fluctuation action}}

The expansion of the action
\begin{equation}
S=-\int_{0}^{\beta}d\tau \int d\mathbf{x}\int \frac{d\mathbf{k}}{\left(
2\pi \right)  ^{3}}\left \{  \frac{1}{\beta}\log \left[  2\cosh \left(
\beta \tilde{E}_{\mathbf{k,Q}}^{\left(  \theta \right)  }\right)  +2\cosh \left(
\beta \tilde{\zeta}_{\mathbf{k,Q}}^{\left(  \theta \right)  }\right)  \right]
-\tilde{\xi}_{\mathbf{k,Q}}^{\left(  \theta \right)  }\right \}  -\beta
V\frac{\Delta^{2}}{g}, \label{A - Actie}%
\end{equation}
up to order $\delta \theta_{\mathbf{x},\tau}^{2}$ is done by first expanding
the action to second order in $\mu_{\mathbf{Q}}^{\left(  \theta \right)  }$ and
$\zeta_{\mathbf{k}}^{\left(  \theta \right)  }$, which are given by (\ref{mu})
and (\ref{zeta}) respectively. After expansion of the integrand of
(\ref{A - Actie}), the action can be divided into a saddle-point contribution
(zeroth order in $\delta \theta_{\mathbf{x},\tau}$):%
\begin{equation}
S_{sp}=-\int_{0}^{\beta}d\tau \int d\mathbf{x}\int \frac{d\mathbf{k}}{\left(
2\pi \right)  ^{3}}\left(  \frac{1}{\beta}\log \left[  2\cosh \left(  \beta
E_{\mathbf{k,Q}}\right)  +2\cosh \left(  \beta \zeta_{\mathbf{k,Q}}\right)
\right]  -\xi_{\mathbf{k,Q}}\right)  -\beta V\frac{\left \vert \Delta
\right \vert ^{2}}{g},
\end{equation}
and a fluctuation contribution (terms of higher order in $\delta
\theta_{\mathbf{x},\tau}$ and its derivatives):%
\begin{align}
S_{fl}  &  =-\int_{0}^{\beta}d\tau \int d\mathbf{x}\int \frac{d\mathbf{k}%
}{\left(  2\pi \right)  ^{3}}\left[  \left(  1-\frac{\xi_{\mathbf{k,Q}}%
}{E_{\mathbf{k,Q}}}X\left(  E_{\mathbf{k,Q}}\right)  \right)  \mu_{\mathbf{Q}%
}^{\left(  \theta \right)  }\right. \nonumber \\
&  +\left[  X\left(  \zeta_{\mathbf{k,Q}}\right)  \right]  \zeta_{\mathbf{k}%
}^{\left(  \theta \right)  }+\left(  \left(  \dfrac{\xi_{\mathbf{k,Q}}%
}{E_{\mathbf{k,Q}}}\right)  ^{2}Y\left(  E_{\mathbf{k,Q}}\right)  +X\left(
E_{\mathbf{k,Q}}\right)  \dfrac{\Delta^{2}}{E_{\mathbf{k,Q}}^{3}}\right)
\frac{\left(  \mu_{\mathbf{Q}}^{\left(  \theta \right)  }\right)  ^{2}}%
{2}\nonumber \\
&  \left.  +Y\left(  E_{\mathbf{k,Q}}\right)  \frac{\left(  \zeta_{\mathbf{k}%
}^{\left(  \theta \right)  }\right)  ^{2}}{2}+2\tilde{Y}\left(  E_{\mathbf{k,Q}%
}\right)  \left(  \dfrac{\xi_{\mathbf{k,Q}}}{E_{\mathbf{k,Q}}}\right)
\frac{\mu_{\mathbf{Q}}^{\left(  \theta \right)  }\zeta_{\mathbf{k}}^{\left(
\theta \right)  }}{2}\right]  ,
\end{align}
where the definitions in (\ref{X,Y,Y_tilde}) were used. The saddle-point
action corresponds to the result of \cite{Devreese Klimin Tempere}. Now the
fluctuation action can be simplified further. Firstly, only the terms up to
order $q^{4}$ are taken into account. Secondly, some terms may vanish due to
e.g. boundary conditions. Now we briefly point out the main simplifications
which can be performed. In the integral over $\mu_{\mathbf{Q}}^{\left(
\theta \right)  }$, the first two terms (we use the order of terms as given in
(\ref{X,Y,Y_tilde})) vanish because of%
\begin{equation}
\int dx~\frac{\partial \delta \theta_{\mathbf{x},\tau}}{\partial x}Q_{x}%
=Q_{x}\left(  \delta \theta_{\mathbf{+\infty},\tau}-\delta \theta
_{\mathbf{-\infty},\tau}\right)  =0,
\label{boundary condition infinity and zero}%
\end{equation}
and the boundary condition for bosons%
\begin{equation}
\int_{0}^{\beta}d\tau \dfrac{\partial \delta \theta_{\mathbf{x},\tau}}%
{\partial \tau}=\delta \theta_{\mathbf{x},\beta}-\delta \theta_{\mathbf{x},0}=0,
\end{equation}
respectively. In the integral over~$\left(  \mu_{\mathbf{Q}}^{\left(
\theta \right)  }\right)  ^{2}$, three terms vanish because they are of other
$\delta \theta_{\mathbf{x},\tau}^{3}$ or higher. In the integral over
$\zeta_{\mathbf{k}}^{\left(  \theta \right)  }$, the first term (again
respecting the order as given in (\ref{X,Y,Y_tilde})) vanishes for the same
reason as (\ref{boundary condition infinity and zero}), while the second term
is zero because of another boundary condition%
\begin{equation}
\int d\mathbf{x}~\nabla_{\mathbf{x}}^{2}\left(  \delta \theta_{\mathbf{x},\tau
}\right)  =\left.  \nabla_{\mathbf{x}}\left(  \delta \theta_{\mathbf{x},\tau
}\right)  \right \vert _{-\infty}^{+\infty}=0.
\end{equation}
Furthermore, in the integral over $\mu_{\mathbf{Q}}^{\left(  \theta \right)
}\zeta_{\mathbf{k}}^{\left(  \theta \right)  }$, two terms again vanish because
they are of order $\delta \theta_{\mathbf{x},\tau}^{3}$ or higher. Finally, the
term%
\begin{equation}
\dfrac{1}{4}\sum_{k}\int_{0}^{\beta}d\tau \int d\mathbf{x}~\smallskip \tilde
{Y}\left(  E_{\mathbf{k,Q}}\right)  \left(  \dfrac{\xi_{\mathbf{k,Q}}%
}{E_{\mathbf{k,Q}}}\right)  \dfrac{\partial \delta \theta_{\mathbf{x},\tau}%
}{\partial \tau}\nabla_{\mathbf{x}}^{2}\left(  \delta \theta_{\mathbf{x},\tau
}\right)  \label{Term (1)}%
\end{equation}
equals zero. Using Green's first identity, the term (\ref{Term (1)}) becomes%
\begin{align}
\int_{0}^{\beta}d\tau \int d\mathbf{x}~\dfrac{\partial \delta \theta
_{\mathbf{x},\tau}}{\partial \tau}\nabla_{\mathbf{x}}^{2}\left(  \delta
\theta_{\mathbf{x},\tau}\right)   &  =-\int_{0}^{\beta}d\tau \int
d\mathbf{x}~\nabla \left(  \dfrac{\partial \delta \theta_{\mathbf{x},\tau}%
}{\partial \tau}\right)  .\nabla_{\mathbf{x}}\left(  \delta \theta
_{\mathbf{x},\tau}\right) \nonumber \\
&  =-\int_{0}^{\beta}d\tau \int d\mathbf{x}~\dfrac{\partial \left[
\nabla_{\mathbf{x}}\left(  \delta \theta_{\mathbf{x},\tau}\right)  \right]
}{\partial \tau}.\nabla_{\mathbf{x}}\left(  \delta \theta_{\mathbf{x},\tau
}\right)  ,
\end{align}
because $\int \left(  \partial \delta \theta/\partial \tau~\nabla \delta
\theta \right)  .d\mathbf{S}=0$. Subsequently, we can write%
\begin{equation}
-\int_{0}^{\beta}d\tau \int d\mathbf{x}~\dfrac{\partial \left[  \nabla
_{\mathbf{x}}\left(  \delta \theta_{\mathbf{x},\tau}\right)  \right]
}{\partial \tau}.\nabla_{\mathbf{x}}\left(  \delta \theta_{\mathbf{x},\tau
}\right)  =-\int_{0}^{\beta}d\tau \int d\mathbf{x}~\frac{1}{2}\dfrac
{\partial \left \{  \left[  \nabla_{\mathbf{x}}\left(  \delta \theta
_{\mathbf{x},\tau}\right)  \right]  ^{2}\right \}  }{\partial \tau}=0, \label{3}%
\end{equation}
again because of boundary conditions. Putting it all together, the fluctuation
action becomes equal to (\ref{Fluctuation action (2)}).

\end{document}